\documentclass[a4paper]{spie}  

\usepackage[]{graphicx}
\usepackage{url}
\usepackage{ifpdf}
\ifpdf
  \usepackage[pdftex]{hyperref}
\else
  \usepackage[ps2pdf]{hyperref}
\fi
\usepackage{amssymb}
\usepackage{amsmath}
\usepackage{bbold}
\usepackage{subfigure}

\newcommand{\eqn}[1]{(#1)}

\newcommand{\fig}[1]{Fig.~#1}

\newcommand{\sectn}[1]{Sec.~#1}

\newcommand{\ie}{\mbox{\it i.e.}}

\newcommand{\spcend}{\ensuremath{\:}}
\newcommand{\img}{\ensuremath{{\rm i}}}
\newcommand{\cconj}{\ensuremath{\ast}} 
\newcommand{\reals}{\ensuremath{\mathbb{R}}}

\newcommand{\integers}{\ensuremath{\mathbb{Z}}}
\newcommand{\naturals}{\ensuremath{\mathbb{N}}}
\newcommand{\complex}{\ensuremath{\mathbb{C}}}
\newcommand{\ltwo}{\ensuremath{\mathrm{L}^2}}
\newcommand{\sphere}{\ensuremath{{\mathrm{S}^2}}}
\newcommand{\sothree}{\ensuremath{{\mathrm{SO}(3)}}}
\newcommand{\torus}{\ensuremath{{\mathrm{T}^2}}}

\newcommand{\vect}[1]{\ensuremath{\mbox{\boldmath ${#1}$}}}

\newcommand{\dx}{\ensuremath{\mathrm{\,d}}}
\newcommand{\dmu}[1]{\ensuremath{\dx \Omega(#1)}}
\newcommand{\dmun}{\ensuremath{\dx \Omega}}

\newcommand{\innerp}[2]{\ensuremath{\langle {#1},\: {#2} \rangle}}

\newcommand{\saa}{\ensuremath{\theta}}
\newcommand{\sab}{\ensuremath{\varphi}}
\newcommand{\sas}{\ensuremath{\saa, \sab}}

\newcommand{\euls}{\ensuremath{\eula, \eulb, \eulc}}
\newcommand{\eula}{\ensuremath{\alpha}}
\newcommand{\eulb}{\ensuremath{\beta}}
\newcommand{\eulc}{\ensuremath{\gamma}}
\newcommand{\el}{\ensuremath{\ell}}
\newcommand{\m}{\ensuremath{m}}
\newcommand{\n}{\ensuremath{n}}
\newcommand{\spin}{\ensuremath{s}}

\newcommand{\elmax}{\ensuremath{{L}}}
\newcommand{\p}{\ensuremath{^\prime}}

\newcommand{\kron}[2]{\ensuremath{\delta_{{#1}{#2}}}}

\renewcommand{\exp}[1]{\ensuremath{{\rm e}^{#1}}}
\newcommand{\shfarg}[3]{\ensuremath{Y_{#1#2}({#3})}}
\newcommand{\shfargc}[3]{\ensuremath{Y_{#1#2}^\cconj({#3})}}

\newcommand{\sshfarg}[4]{\ensuremath{{{}_{#4} Y_{#1#2}({#3})}}}

\newcommand{\shf}[2]{\ensuremath{Y_{#1#2}}}

\newcommand{\shc}[3]{\ensuremath{{#1}_{{#2}{#3}}}}

\newcommand{\leg}[2]{\ensuremath{P_{{#1}}({#2})}}
\newcommand{\aleg}[3]{\ensuremath{P_{#1}^{#2}({#3})}}

\newcommand{\dmatbig}{\ensuremath{D}}

\newcommand{\dmatsmall}{\ensuremath{d}}
\newcommand{\dlmn}{\ensuremath{ \dmatsmall_{\m\n}^{\el} }}

\newcommand{\dlmnb}{\ensuremath{ \dmatsmall_{\m\n}^{\el}(\eulb) }}

\newcommand{\dlmnhalfpi}[3]{\ensuremath{ \Delta_{{#2}{#3}}^{#1} }}

\newcommand{\dlmnhalfpim}{\ensuremath{ \Delta_{{\m\p}{\m}}^{\el} }}

\newcommand{\f}{\ensuremath{f}}
\newcommand{\fs}{\ensuremath{{}_\spin f}}

\newcommand{\flm}{\ensuremath{\shc{\f}{\el}{\m}}}
\newcommand{\fslm}{\ensuremath{\shc{\fs}{\el}{\m}}}

\newcommand{\elmfact}{\ensuremath{\frac{(\el-\m)!}{(\el+\m)!}}}

\newcommand{\sumlm}{\ensuremath{\sum_{\el=0}^{\infty} \sum_{\m=-\el}^\el}}

\newcommand{\summ}{\ensuremath{\sum_{\m=-\el}^\el}}

\newcommand{\nl}{\ensuremath{\sqrt{\frac{2\el+1}{4\pi}}}}

\newcommand{\G}[3]{\ensuremath{{{}_{#1} G_{{#2} {#3}}}}}

\newcommand{\rG}[3]{\ensuremath{{{}_{#1} \tilde{G}_{{#2} {#3}}}}}
\newcommand{\Gsm}{\ensuremath{\G{\spin}{\m}{}}}
\newcommand{\Gsmm}{\ensuremath{\G{\spin}{\m}{\m\p}}}
\newcommand{\Gsmt}{\ensuremath{\G{\spin}{\m}{}(\saa)}}

\newcommand{\Gsmti}{\ensuremath{\G{\spin}{\m}{}({\saaiang})}}

\newcommand{\rGsmti}{\ensuremath{\rG{\spin}{\m}{}({\saaiang})}}
\newcommand{\F}[3]{\ensuremath{{{}_{#1} F_{{#2} {#3}}}}}

\newcommand{\Fsmm}{\ensuremath{\F{\spin}{\m}{\m\p}}}

\newcommand{\Fsmmp}{\ensuremath{\F{\spin}{\m}{\m{\p}{\p}}}}

\newcommand{\intsaa}{\ensuremath{\int_0^{\pi} \dx \saa \sin \saa}}
\newcommand{\intsab}{\ensuremath{\int_0^{2\pi} \dx \sab}}
\newcommand{\saai}{\ensuremath{t}}
\newcommand{\sabi}{\ensuremath{p}}
\newcommand{\saaiang}{\ensuremath{\saa_\saai}}
\newcommand{\sabiang}{\ensuremath{\sab_\sabi}}

\newcommand{\saisang}{\ensuremath{\saaiang,\sabiang}}
\newcommand{\sumsaai}{\ensuremath{\sum_{\saai=-(\elmax-1)}^{\elmax-1}}}
\newcommand{\sumsabi}{\ensuremath{\sum_{\sabi=-(\elmax-1)}^{\elmax-1}}}

\newcommand{\summptrunc}{\ensuremath{\sum_{\m\p=-(\elmax-1)}^{\elmax-1}}}
\newcommand{\qweight}{\ensuremath{q}}
\newcommand{\qweightdh}{\ensuremath{\qweight_{\rm DH}}}
\newcommand{\qweightmw}{\ensuremath{\qweight_{\rm MW}}}
\newcommand{\weighttrans}{\ensuremath{v}}

\newcommand{\N}{\ensuremath{{N}}}

\newcommand{\Ndh}{\ensuremath{{N_{\rm DH}}}}

\newcommand{\Nmw}{\ensuremath{{N_{\rm MW}}}}

\newcommand{\weight}{\ensuremath{w}}
\newcommand{\order}{\ensuremath{\mathcal{O}}}

\newcommand{\nmeas}{\ensuremath{M}}

\newcommand{\sparmat}{\ensuremath{\Psi}}
\newcommand{\sensmat}{\ensuremath{\Phi}}

\renewcommand{\eqn}[1]{Eqn.~(#1)}

\renewcommand{\sparmat}{\ensuremath{\Lambda}}
\renewcommand{\spin}{\ensuremath{\mathbf{s}}}
\renewcommand{\spin}{{}}

\title{Sampling theorems and compressive sensing on the sphere}

\author{Jason D. McEwen\supit{a}, 
  Gilles~Puy\supit{a},
  Jean-Philippe Thiran\supit{a},
  Pierre Vandergheynst\supit{a},\\
  Dimitri Van De Ville\supit{b,c}
  and Yves~Wiaux\supit{a,b,c}
\skiplinehalf
\supit{a} Institute of Electrical Engineering, Ecole Polytechnique F{\'e}d{\'e}rale de Lausanne (EPFL), \\ Lausanne 1015, Switzerland; \\
\supit{b} Institute of Bioengineering, Ecole Polytechnique F{\'e}d{\'e}rale de Lausanne (EPFL), \\ Lausanne 1015, Switzerland; \\
\supit{c} Department of Radiology and Medical Informatics, University of Geneva (UniGE), \\ Geneva 1211,
  Switzerland
}

\authorinfo{Further author information:\\JDM: E-mail: {\tt
    mcewen@mrao.cam.ac.uk}, URL: \url{http://www.jasonmcewen.org}}

\begin{document} 
\maketitle 


\begin{abstract}
  We discuss a novel sampling theorem on the sphere developed by
  McEwen \& Wiaux recently through an association between the sphere
  and the torus.  To represent a band-limited signal exactly, this new
  sampling theorem requires less than half the number of samples of
  other equiangular sampling theorems on the sphere, such as the
  canonical Driscoll \& Healy sampling theorem. A reduction in the
  number of samples required to represent a band-limited signal on the
  sphere has important implications for compressive sensing,
  both in terms of the dimensionality and sparsity of signals. We
  illustrate the impact of this property with an inpainting problem on
  the sphere, where we show superior reconstruction performance when
  adopting the new sampling theorem.
\end{abstract}

\keywords{Sphere, spherical harmonics, sampling theorem, compressive sensing.}

\section{INTRODUCTION} 

The fast Fourier transform\cite{cooley:1965} (FFT) is arguably the
most important and widely used numerical algorithm of our era,
rendering the frequency content of signals accessible.  Moreover,
Shannon's sampling theorem\cite{shannon:1949} states that all of the
information content of a band-limited continuous signal can be
captured through a finite number of samples.  Typically, standard
Fourier analyses are performed in Euclidean space where Shannon's
theory holds and where FFTs are directly applicable.  However, in many
applications data are observed on non-Euclidean manifolds, such as the
sphere.  Fourier analysis is performed on the sphere in the basis of
spherical harmonics, which are the eigenfunctions of the spherical
Laplacian operator and form the canonical orthonormal basis on the
sphere.  Sampling theorems and fast algorithms to perform spherical
harmonic analyses exist but the field is much less mature that its
elder Euclidean sibling.

A novel sampling theorem on the sphere has been developed recently by
two of the authors of this article \cite{mcewen:fssht} (hereafter
referred to as the MW sampling theorem).  From an information
theoretic viewpoint, the fundamental property of any sampling theorem
is the number of samples required to capture all of the information
content of a band-limited signal.  To represent exactly a signal on
the sphere band-limited at \elmax, all sampling theorems on the sphere
require $\order(\elmax^2)$ samples. However, the MW sampling theorem
requires $\sim2\elmax^2$ samples only, less than half of the
$\sim4\elmax^2$ samples of other equiangular sampling theorems on the
sphere, such as the canonical Driscoll \& Healy sampling
theorem\cite{driscoll:1994} (hereafter referred to as the DH sampling
theorem).  Not only is the MW sampling theorem of theoretical
interest, particularly regarding the information content of signals on
the sphere, but it also has important practical implications in the
emerging field of compressive sensing.

The theory of compressive sensing states that it is possible to
acquire sparse or compressible signals with fewer samples than
standard sampling theorems would suggest
\cite{candes:2006a,donoho:2006}.  In these settings, the ratio of the
required number of measurements to the dimensionality of the signal
scales linearly with its sparsity \cite{candes:2006a}.  The more
efficient sampling of the MW sampling theorem reduces the
dimensionality of the signal in the spatial domain, thereby improving
the performance of compressive sensing reconstruction on the sphere
when compared to alternative sampling theorems.\cite{mcewen:css2}
Furthermore, for sparsity priors defined in the spatial domain, such
as signals sparse in the magnitude of their gradient, sparsity is also
directly related to the sampling of the signal.  For this class of
signals, an additional enhancement in compressive sensing
reconstruction performance is thus achieved when adopting the MW
sampling theorem. \cite{mcewen:css2}

In this article we first review sampling theorems on the sphere in
\sectn{\ref{sec:sampling_theorems}}, focussing on the MW and DH
sampling theorems.  In \sectn{\ref{sec:compressive_sensing}} we
discuss the superior performance achieved when solving compressive
sensing problems on the sphere using the MW sampling theorem, as
opposed to the DH sampling theorem.  We illustrate our arguments with
an inpainting problem on the sphere, where we adopt the prior
assumption that the signal to be recovered is sparse in the magnitude
of its gradient.  Simulations are performed, verifying our theoretical
arguments.  Finally, concluding remarks are made in
\sectn{\ref{sec:conclusions}}.

\section{SAMPLING THEOREMS ON THE SPHERE} 
\label{sec:sampling_theorems}

Sampling theorems on the sphere state that all of the information
contained in a band-limited signal may be represented by a finite set
of samples in the spatial domain.  On the sphere, unlike Euclidean
space, the number of samples required in the harmonic and spatial
domains differ, with different sampling theorems on the sphere
requiring a different number of samples in the spatial domain.
For an equiangular sampling of the sphere, the DH sampling theorem has
become the standard, while the MW sampling theorem has emerged only
recently.\footnote{Fast algorithms have been developed to compute the
  forward and inverse transforms rapidly for both the DH
  \cite{driscoll:1994,healy:2003} and MW \cite{mcewen:fssht} sampling
  theorems; these algorithms are essential to facilitate the
  application of these sampling theorems at high band-limits.}  The MW
sampling theorem achieves a more efficient sampling of the sphere,
with a reduction by a factor of two in the number of samples required
to represent a band-limited signal.\footnote{Gauss-Legendre (GL)
  quadrature can also be used to construct an efficient sampling
  theorem on the sphere, with \mbox{$\sim 2 \elmax^2$} samples
  \cite{mcewen:fssht}.  The MW sampling theorem nevertheless requires
  fewer samples and so remains more efficient, especially at low
  band-limits.  Furthermore, it is not so straightforward to define
  norms describing spatial priors on the GL grid since it is not
  equiangular.  Finally, algorithms implementing the GL sampling
  theorem have been shown to be limited to lower band-limits and less
  accurate than the algorithms implementing the MW sampling theorem
  \cite{mcewen:fssht}. Consequently, we do not focus on the GL
  sampling theorem any further in this article.}  In this section we
outline the harmonic structure of the sphere in the continuous
setting, before reviewing concisely the DH and MW sampling theorems.

\subsection{Harmonic Analysis on the Sphere} 

We consider the space of square integrable functions on the sphere $\ltwo(\sphere)$, with the inner product of $f,g\in\ltwo(\sphere)$ defined by
\begin{equation}
\innerp{f}{g} = \int_\sphere \dmu{\sas} \: f(\sas) \: g^\cconj(\sas) 
\spcend ,
\end{equation}
where $\dmu{\sas} = \sin \saa \dx \saa \dx \sab$ is the usual invariant measure on the sphere and $(\sas)$ define spherical coordinates with colatitude $\saa \in [0,\pi]$ and longitude $\sab \in [0,2\pi)$.   Complex conjugation is denoted by the superscript ${}^\cconj$.  

The spherical harmonic functions form the canonical orthogonal basis for the space of $\ltwo(\sphere)$ functions on the sphere and are defined by
\begin{equation}
\shfarg{\el}{\m}{\sas} = 
\sqrt{\frac{2\el+1}{4\pi} \elmfact} \:
\aleg{\el}{\m}{\cos\saa} \:
\exp{\img \m \sab} 
\spcend ,
\end{equation}
for natural $\el\in\naturals$ and integer $\m\in\integers$, $|\m|\leq\el$, where $\aleg{\el}{\m}{x}$ are the associated Legendre functions. 
We adopt the Condon-Shortley phase convention, with the $(-1)^\m$ phase factor included in the definition of the associated Legendre functions.
The orthogonality and completeness relations for the spherical harmonics read
$
\innerp{\shf{\el}{\m}}{\shf{\el\p}{\m\p}}
= 
\kron{\el}{\el\p}
\kron{\m}{\m\p}
$
and
\begin{equation}
\sumlm
\shfarg{\el}{\m}{\saa,\sab} \:
\shfargc{\el}{\m}{\saa\p,\sab\p} 
=
\delta(\cos\saa - \cos\saa\p) \:
\delta(\sab - \sab\p)
\end{equation}
respectively, where $\kron{i}{j}$ is the Kronecker delta symbol and $\delta(x)$ is the Dirac delta function.

Due to the orthogonality and completeness of the spherical harmonics, any square integrable function on the sphere $\f \in \ltwo(\sphere)$ may be represented by its spherical harmonic expansion
\begin{equation}
\label{eqn:sht_inverse}
\f(\sas) = 
\sum_{\el=0}^\infty
\sum_{\m=-\el}^\el
\flm \:
\shfarg{\el}{\m}{\sas}
\spcend ,
\end{equation}
where the spherical harmonic coefficients are given by the usual projection onto each basis function:
\begin{equation}
\label{eqn:sht_forward}
\shc{\f}{\el}{\m} =
\innerp{f}{\shf{\el}{\m}}
=
\int_\sphere \dmu{\sas} \: f(\sas) \: \shfargc{\el}{\m}{\sas}
\spcend .
\end{equation}
Throughout, we consider signals on the sphere band-limited at
$\elmax$, that is signals such that $\shc{f}{\el}{\m}=0$, $\forall
\el\geq\elmax$.  In this case the summation over \el\ in
\eqn{\ref{eqn:sht_inverse}} may be truncated to $\elmax-1$.  We also
adopt the convention that harmonic coefficients are defined to be zero
for $|\m|>\el$ (which enforces the contraint $|\m| \leq \el$ when
summations are interchanged).

The forward and inverse spherical harmonic transforms, given by
\eqn{\ref{eqn:sht_forward}} and \eqn{\ref{eqn:sht_inverse}}
respectively, are exact in the continuous setting.  A sampling theorem
on the sphere states how to sample a band-limited function $\f(\sas)$
at a finite number of locations, such that all of the information
content of the continuous function is captured.  Since the frequency
domain of a function on the sphere is discrete, the spherical harmonic
coefficients describe the continuous function exactly.  A sampling
theorem thus describes how to exactly recover the spherical harmonic
coefficients \flm\ of the continuous function from its samples.
Consequently, sampling theorems effectively encode (often implicitly)
an exact quadrature rule for evaluating the integral of a band-limited
function on the sphere.

\subsection{Driscoll \& Healy Sampling Theorem} 

The DH sampling theorem\cite{driscoll:1994} gives an explicit
quadrature rule for the evaluation of spherical harmonic coefficients: 
\begin{equation}
\label{eqn:dh_quad}
\shc{\f}{\el}{\m} =
\sum_{\saai=0}^{2\elmax-1} \:
\sum_{\sabi=0}^{2\elmax-1} \:
\qweightdh(\saaiang) \: \f(\saisang) \: \shfargc{\el}{\m}{\saisang}
\spcend ,
\end{equation}
where the equiangular sample positions are defined by $\saaiang= \pi
\saai / 2\elmax $, for $\saai = 0, \dotsc, 2\elmax-1$, and $\sabiang =
\pi\sabi / \elmax$, for $\sabi = 0, \dotsc, 2\elmax -1$, giving
$\Ndh=(2\elmax-1)2\elmax + 1 \sim 4\elmax^2$ samples on the
sphere.\footnote{The original DH sampling theorem has been
  revisited\cite{healy:2003}, yielding an alternative formulation with
  only very minor differences and that also requires $\sim4\elmax^2$
  samples.}
The quadrature used to evaluate \eqn{\ref{eqn:sht_forward}} is exact
for a function band-limited at \elmax, with quadrature weights given
implicitly by the solution to
\begin{equation}
\label{eqn:dh_quad_weight_implicit}
\sum_{\saai=0}^{2\elmax-1} \:
\qweightdh(\saaiang) \: \leg{\el}{\cos \saaiang}
= \frac{2\pi}{\elmax} \: \kron{\el}{0}
\spcend ,
\quad \forall \el < 2\elmax
\spcend .
\end{equation}
The quadrature weights satisfying
\eqn{\ref{eqn:dh_quad_weight_implicit}} are given
by
\begin{equation}
\qweightdh(\saaiang) =
\frac{2\pi}{\elmax^2} \:
\sin\saaiang \:
\sum_{k=0}^{\elmax-1} \:
\frac{\sin((2k+1)\saaiang)}{2k+1}
\spcend .
\end{equation}

The exactness of the quadrature rule is proved by considering the
sampling distribution of Dirac delta functions defined by 
\begin{equation}
\label{eqn:dh_sampling_distn}
s(\sas) = 
\sum_{\saai=0}^{2\elmax-1} \:
\sum_{\sabi=0}^{2\elmax-1} \:
\qweightdh(\saaiang) \: 
\delta(\cos\saa - \cos\saaiang) \: \delta(\sab - \sabiang)
\spcend .
\end{equation}
For quadrature weights satisfying
\eqn{\ref{eqn:dh_quad_weight_implicit}}, it can be shown that
$\shc{s}{0}{0}=\sqrt{4\pi}$ and $\shc{s}{\el}{\m}=0$ for
$0<\el<2\elmax$, $\forall \m$.  Consequently, the sampling
distribution may be written
\begin{equation}
\label{eqn:dh_sampling_distn_2}
s(\sas) = 
1 +
\sum_{\el=2\elmax}^{\infty} \:
\summ \:
\shc{s}{\el}{\m} \: \shfarg{\el}{\m}{\sas}
\spcend .
\end{equation}
The harmonic coefficients of the product of the original band-limited
function and the sampling distribution $\f^s = \f \cdot s $ are given by
\begin{equation}
\shc{\f}{\el}{\m}^s =
\sum_{\saai=0}^{2\elmax-1} \:
\sum_{\sabi=0}^{2\elmax-1} \:
\qweightdh(\saaiang) \: \f(\saisang) \: \shfargc{\el}{\m}{\saisang}
\spcend ,
\end{equation}
which follows from \eqn{\ref{eqn:dh_sampling_distn}}.  Notice that
these harmonic coefficients are given by the quadrature rule specified
in \eqn{\ref{eqn:dh_quad}} and it simply remains to prove that the
harmonic coefficients of $\f^s$ agree with those of \f\ for the
harmonic range of interest (\ie\ for $0\leq\el<\elmax$).  Noting
\eqn{\ref{eqn:dh_sampling_distn_2}}, we may write $\f^s(\sas) = 
\f(\sas) + \alpha(\sas)$, where 
\begin{equation}
\alpha(\sas) = 
\sum_{\el=2\elmax}^{\infty} \:
\summ \:
\shc{s}{\el}{\m} \:
\shfarg{\el}{\m}{\sas} \:
\sum_{\el\p=0}^{\elmax-1} \:
\sum_{\m\p=-\el\p}^{\el\p} \:
\shc{\f}{\el\p}{\m\p} \:
\shfarg{\el\p}{\m\p}{\sas} 
\spcend .
\end{equation}
Since the product of two spherical harmonic functions
$\shfarg{\el}{\m}{\sas} \: \shfarg{\el\p}{\m\p}{\sas}$ can be written
as a sum of spherical harmonics with minimum degree $| \el - \el\p| $,
\cite{driscoll:1994} the aliasing error $\alpha(\sas)$ contains
non-zero harmonic content for $\el>\elmax$ only.  Aliasing is
therefore outside of the harmonic range of interest and
$\shc{\f}{\el}{\m}^s = \shc{\f}{\el}{\m}$ for $0\leq\el<\elmax$,
$|m|<\el$, thus proving the exact quadrature rule given by
\eqn{\ref{eqn:dh_quad}}.

\subsection{McEwen \& Wiaux Sampling Theorem} 

The MW sampling theorem \cite{mcewen:fssht} follows by a factoring of
rotations \cite{risbo:1996} and a periodic extension in colatitude
\saa, so that the orthogonality of the complex exponentials over $[0,
2\pi)$ may be exploited.  This approach encodes an implicit quadrature
rule on the sphere, which can then be made explicit.

The spherical harmonics are related to the Wigner functions
through \cite{goldberg:1967}
\begin{equation}
\label{eqn:ssh_wigner}
\sshfarg{\el}{\m}{\sas}{\spin} =
\sqrt{\frac{2\el+1}{4\pi} } \:
\dmatbig_{\m 0}^{\el\:\cconj}(\sab,\saa  ,0)
\spcend ,
\end{equation}
where the Wigner functions form the canonical orthogonal basis on the
rotation group \sothree.  The Wigner functions may be decomposed as
\cite{varshalovich:1989}
\begin{equation}
\label{eqn:d_decomp}
\dmatbig_{\m\n}^{\el}(\euls)
= {\rm e}^{-\img \m\eula} \:
\dmatsmall_{\m\n}^\el(\eulb) \:
{\rm e}^{-\img \n\eulc}
\spcend ,
\end{equation}
where the rotation group is parameterised by the Euler angles
$(\euls)$, with $\eula \in [0,2\pi)$, $\eulb \in [0,\pi]$ and $\eulc
\in [0,2\pi)$.  The Fourier series decomposition of the
\dmatsmall-functions is given by \cite{nikiforov:1991}
\begin{equation}
  \label{eqn:wigner_sum_reln}
  \dlmnb = \img^{\n-\m} \sum_{\m\p=-\el}^\el
  \dlmnhalfpi{\el}{\m\p}{\m} \:
  \dlmnhalfpi{\el}{\m\p}{\n} \:
  \exp{\img \m\p \eulb}
  \spcend ,
\end{equation}
with \mbox{$\dlmnhalfpi{\el}{\m}{\n} = \dlmn (\pi/2)$}.  This
expression follows from a factoring of rotations. \cite{risbo:1996}
The Fourier series representation of \dlmnb\ given by
\eqn{\ref{eqn:wigner_sum_reln}} allows one to write the spherical
harmonic expansion of $\f(\sas)$ in terms of a Fourier series
expansion of the function extended appropriately to the two-torus
\torus.  Noting
\eqn{\ref{eqn:ssh_wigner}} -- \eqn{\ref{eqn:wigner_sum_reln}}, the
forward spherical harmonic transform may be written
\begin{equation}
  \label{eqn:flm_wig}
  \fslm = 
  \img^{\m} \nl
  \summptrunc
  \dlmnhalfpim \:
  \dlmnhalfpi{\el}{\m\p}{0} \:
  \Gsmm
  \spcend ,
\end{equation}
where 
\begin{equation}
  \label{eqn:Gmm}
  \Gsmm = \intsaa \: \Gsmt \: \exp{-\img \m\p \saa}
\end{equation}
and
\begin{equation}
  \label{eqn:Gmt}
  \Gsmt = \intsab \: \fs(\sas) \: \exp{-\img \m \sab}
  \spcend .
\end{equation}
Since \eqn{\ref{eqn:Gmt}} is simply a Fourier transform, the discrete
and continuous orthogonality of the complex exponentials may be
exploited to evaluate this integral exactly by
\begin{equation}
 \Gsmti = \frac{2 \pi}{2\elmax-1} 
 \sumsabi \: 
 \fs(\saisang) \: 
 \exp{-\img \m \sabiang}
 \spcend ,
\end{equation}
where $\sab_\sabi = 2 \pi \sabi/(2\elmax-1)$, for $\sabi =
0,\dotsc,2\elmax-2$, and $\saa_\saai =
\pi(2\saai+1)/(2\elmax-1)$, for $\saai = 0,\dotsc,\elmax-1$,
giving $\Nmw = (\elmax - 1) (2 \elmax - 1) + 1 \sim 2 \elmax^2$
samples on the sphere.
It remains to develop a quadrature rule to evaluate
\eqn{\ref{eqn:Gmm}}.  This is achieved by extending \Gsmt\ to the
domain $\saa \in [0,2\pi)$ through the construction
\begin{align*}
 \rGsmti =  
 &\quad
 \begin{cases}
    \Gsmti \: , & \saai \in \{ 0, 1, \dotsc, \elmax-1 \} \\
   (-1)^{\m} \: 
   \Gsm(\saa_{2\elmax-2-\saai}) \: , & \saai \in \{ \elmax, \dotsc, 2\elmax-2 \}
 \end{cases}
 \spcend ,
\end{align*}
so that \rGsmti\ may be expressed by a Fourier series.  The factor
$(-1)^\m$ is required to ensure that the symmetry in the domain
$[0,2\pi)$ dictated by the inverse transform is preserved.
Substituting the Fourier expansion of \rGsmti\ into
\eqn{\ref{eqn:Gmm}} yields
\begin{equation}
 \Gsmm = 2 \pi \sum_{\m{\p}{\p}=-(\elmax-1)}^{\elmax-1} \: \Fsmmp \: \weight(\m{\p}{\p} - \m\p)
 \spcend ,
 \label{eqn:Gmm_convolution}
\end{equation}
where the weights are given by
\begin{equation}
 \weight(\m\p)
 = \intsaa \: \exp{\img \m\p \saa}
 = 
 \begin{cases}
   \: \pm \img \pi/2, & \m\p=\pm 1\\
   \: 0, & \m\p \text{ odd},\:\m\p\neq\pm1\\
   \: 2/(1-{\m\p}^2), & \m\p \text{ even}
 \end{cases}
 \spcend ,
\end{equation}
with
\begin{equation*}
 \Fsmm =
 \frac{1}{2 \pi(2\elmax-1)} 
 \sumsaai \: \rGsmti \:
 \exp{- \img \m\p \saaiang}
 \spcend .
\end{equation*}
Since the spherical harmonic coefficients \flm\ are recovered exactly,
all of the information content of the function $\f(\sas)$ is captured
in the finite set of samples.

The derivation above effectively gives an implicit quadrature rule for
the exact integration of a band-limited function on the sphere.  This
quadrature rule can be written explicitly as \cite{mcewen:fssht}
\begin{equation}
\label{eqn:quadrature}
\int_\sphere \dmu{\sas} \: \fs(\sas) = 
\sum_{\saai=0}^{\elmax-1} \:
\sum_{\sabi=0}^{2\elmax-2} \:
\qweightmw(\saaiang) \:
\fs(\saisang)
\spcend ,
\end{equation}
where the quadrature weights are defined by
\begin{equation}
\qweightmw(\saaiang) = 
\frac{2\pi}{2\elmax-1} 
\Bigl[ 
\weighttrans(\saaiang) 
+ (1 - \kron{t,}{\elmax-1}) \: 
\weighttrans(\saa_{2\elmax-2-\saai})
\Bigr]
\spcend ,
\end{equation}
and where $\weighttrans(\saaiang)$ is the inverse discrete Fourier transform of the reflected weights $\weight(-\m\p)$:
\begin{equation}
\weighttrans(\saaiang) =
\frac{1}{2\elmax-1} \: \sum_{\m\p=-(\elmax-1)}^{\elmax-1} \:
\weight(-\m\p) \: \exp{\img \m\p \saaiang}
\spcend .
\end{equation}

\section{COMPRESSIVE SENSING ON THE SPHERE} 
\label{sec:compressive_sensing}

Compressive sensing on the sphere has been studied recently for
signals sparse in harmonic space or in a redundant set of overcomplete
dictionaries. \cite{abrial:2007,rauhut:2011} However, many natural
signals are sparse in measures defined in the spatial domain, such as
in the magnitude of their gradient.  A more efficient sampling of a
band-limited signal on the sphere, as afforded by the MW sampling
theorem, improves both the dimensionality and sparsity of the signal
in the spatial domain, which has been shown to improve the quality of
compressive sampling reconstruction.\cite{mcewen:css2} We review this
very recent work, discussing the impact of efficient sampling on the
sphere in the context of a total variation (TV) inpainting problem,
after first defining the discrete TV norm on the sphere.

\subsection{TV Norm on the Sphere} 

The continuous TV norm on the sphere is defined by 
\begin{equation*}
\| \f \|_{\rm TV} 
=
\int_\sphere \dmun \:
| \nabla \f |
\spcend ,
\end{equation*}
where the magnitude of the gradient of the signal $\f$ is given by 
\begin{equation*}
| \: \nabla \f  \:| =
\sqrt{
\Biggl ( \frac{\partial \f}{\partial \saa} \Biggr)^2
+
\frac{1}{\sin^2\saa}\Biggl (  \frac{\partial \f}{\partial \sab} \Biggr )^2
}
\spcend .
\end{equation*}
In practice, however, one must consider the TV norm of
the sampled signal, where the samples of $\f(\sas)$ are denoted by the
concatenated vector $\vect{x}\in\complex^{\N}$, where \N\ is the
number of samples on the sphere of the chosen sampling theorem
(hereafter harmonic coefficients \flm\ are also represented by a
concatenated vector, denoted $\vect{\hat{x}}\in\complex^{\elmax^2}$).
A discrete TV norm on the sphere is defined by approximating the
continuous norm in the context of either the DH or MW sampling
theorem.  The integral of the continuous TV norm can be approximated
using the quadrature rule corresponding to the sampling theorem on the
sphere adopted:
\begin{equation}
\label{eqn:tvnorm_cts_approx}
\| \f \|_{\rm TV} 
\simeq
\sum_{\saai=0}^{\N_{\saa}-1} \:
\sum_{\sabi=0}^{\N_{\sab}-1} \:
| \nabla \f | \:
\qweight(\saa_\saai)
\spcend ,
\end{equation}
where the number of samples in $(\sas)$, given by $\N_\saa$ and
$\N_\sab$ respectively, and the quadrature weights
$\qweight(\saa_\saai)$ depend on the choice of sampling theorem.  If
$| \nabla \f |$ were band-limited, then
\eqn{\ref{eqn:tvnorm_cts_approx}} would be exact.  Although this is
not likely to be the case, \eqn{\ref{eqn:tvnorm_cts_approx}} is
nevertheless a reasonable approximation of the continuous TV norm.
The magnitude of the gradient $|\nabla \f|$ can be approximated from
the samples $\vect{x}$ using finite differences, to give a discrete TV
norm on the sphere that approximates the continuous norm closely: $\|
\vect{x} \|_{\rm TV} \simeq \| \f \|_{\rm TV}$.  Notice that the
inclusion of the quadrature weights $\qweight(\saa_\saai)$ regularises
the $\sin{\saa}$ term that arises from the definition of the gradient
on the sphere, eliminating numerical instabilities that would
otherwise occur.

\subsection{TV Inpainting on the Sphere} 

We illustrate the impact of the number of samples of the DH and MW
sampling theorems on compressive sensing reconstruction with an
inpainting problem, where measurements are made in the spatial domain.
A test signal sparse in its gradient is constructed from a binary
Earth map, smoothed to give a signal band-limited at $\elmax=32$ (see
\fig{\ref{fig:spheres}~(a)}).\footnote{The original Earth topography
  data are taken from the Earth Gravitational Model (EGM2008) publicly
  released by the U.S. National Geospatial-Intelligence Agency (NGA)
  EGM Development Team.  These data were downloaded and extracted
  using the tools available from Frederik Simons' webpage:
  \url{http://www.princeton.edu/geosciences/people/simons/}.}  The
real inpainting problem 
$\vect{y} = 
\sensmat \vect{x} + \vect{n}$
is considered, 
where $\nmeas$ noisy measurements $\vect{y}\in\reals^{\nmeas}$ of the
signal on the sphere $\vect{x}\in\reals^{\N}$ are made.  The measurement
operator $\sensmat\in\reals^{\nmeas \times \N}$ represents a random
masking of the signal. The noise $\vect{n}\in\reals^\nmeas$ is assumed
to be independent and identically distributed Gaussian noise,
with zero mean.

\newlength{\sphereplotwidth}
\setlength{\sphereplotwidth}{54mm}

\begin{figure}
\centering
\subfigure[Ground truth]{\includegraphics[clip=,viewport=2 2 438 223,width=\sphereplotwidth]{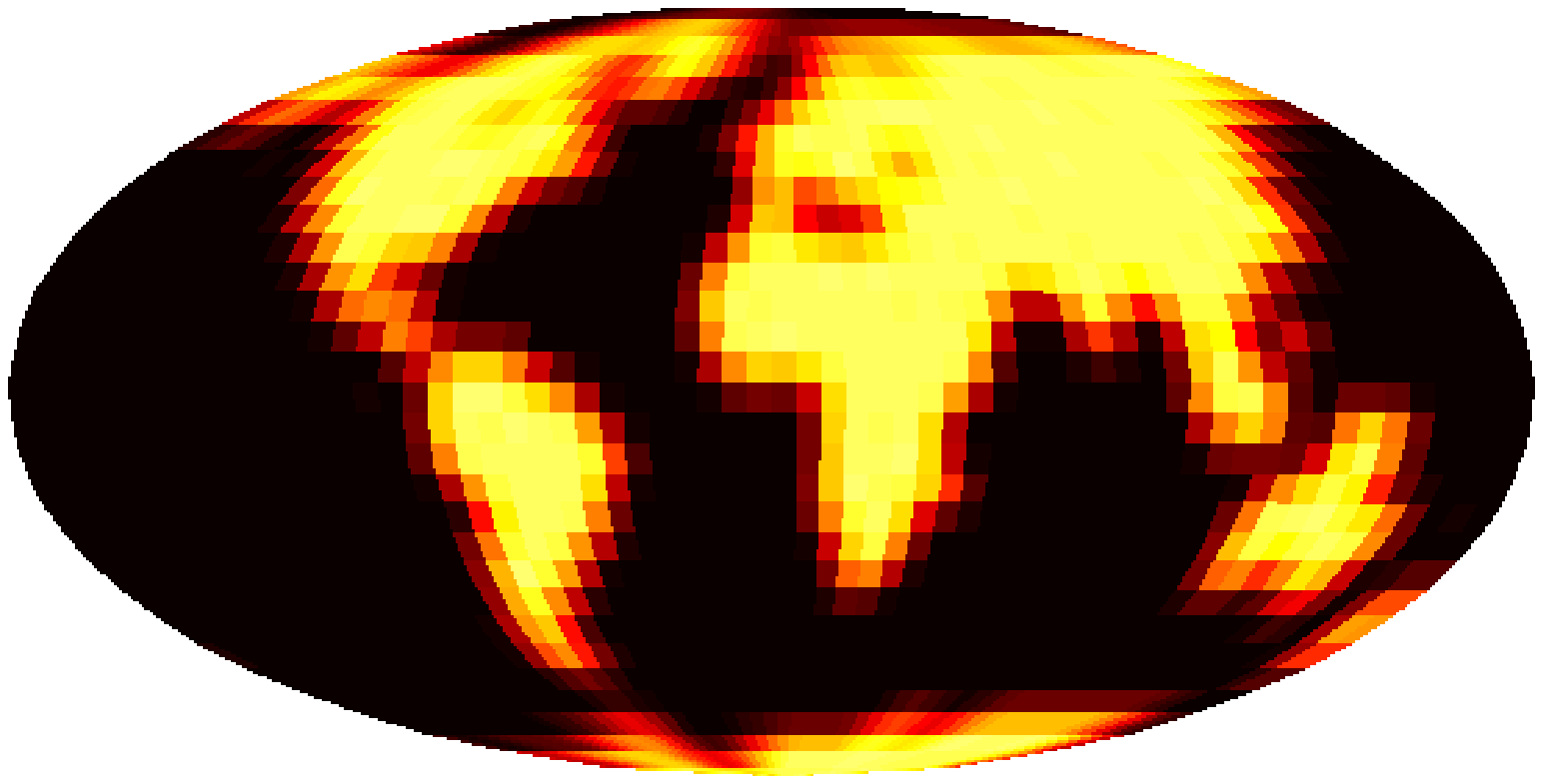}}\quad
\subfigure[DH reconstruction]{\includegraphics[clip=,viewport=2 2 438 223,width=\sphereplotwidth]{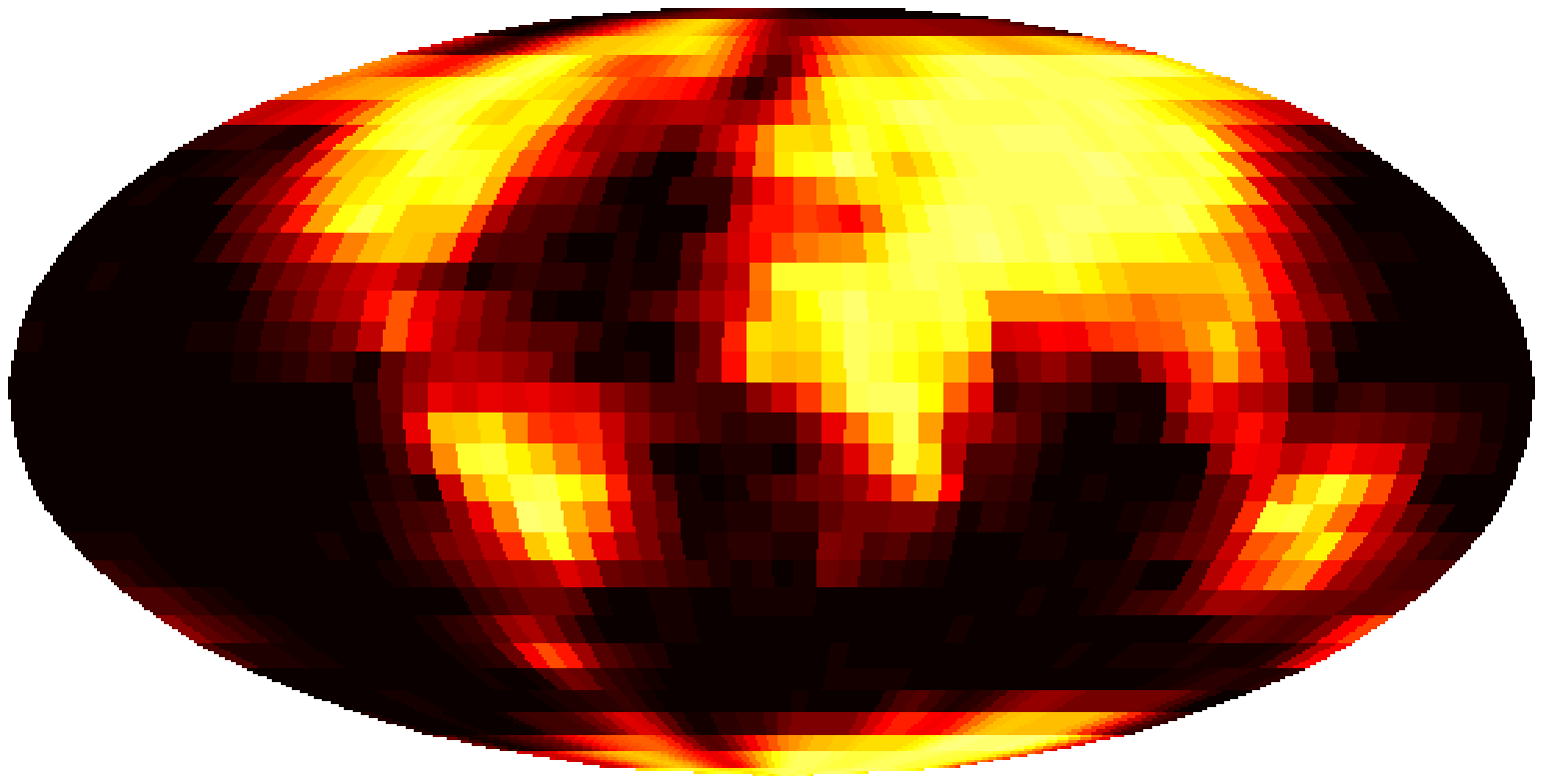}} \quad
\subfigure[MW reconstruction]{\includegraphics[clip=,viewport=2 2 438 223,width=\sphereplotwidth]{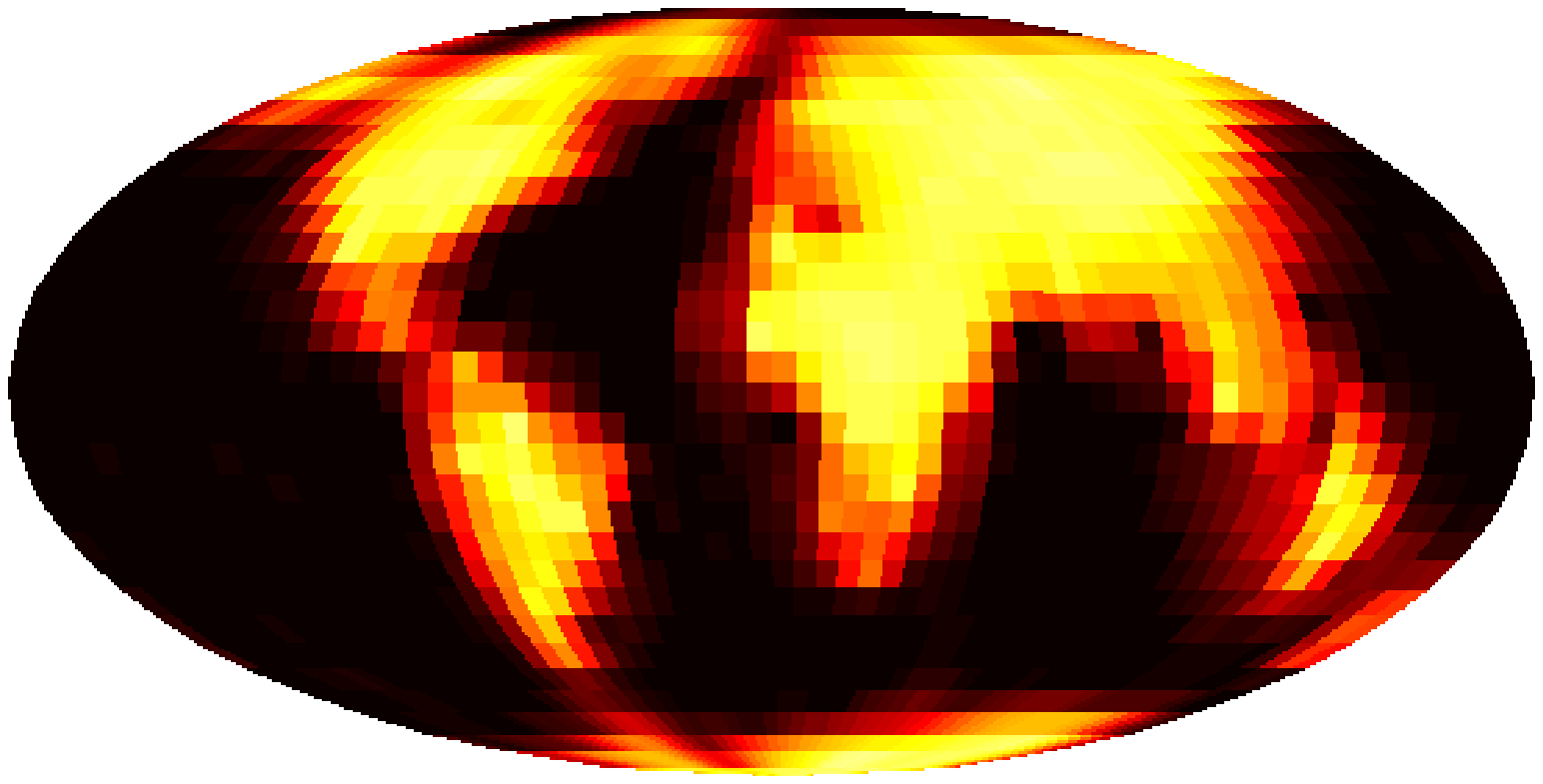}}
\caption{Earth topographic data reconstructed in the harmonic domain for $\nmeas/\elmax^2 = 1/2$}
\label{fig:spheres}
\end{figure}

The TV inpainting problem is first solved directly on the sphere:
\begin{equation}
\label{eqn:recon_spatial}
\vect{x}^\star =
\underset{\vect{x}}{\arg \min} \:
\| \vect{x} \|_{\rm TV} \:\: \mbox{such that} \:\:
\| \vect{y} - \sensmat \vect{x}\|_2 \leq \epsilon
\spcend ,
\end{equation}
where the bound $\epsilon$ is related to a residual noise level
estimator.  By adopting the MW sampling theorem in place of the DH
sampling theorem, the dimensionality and sparsity of the signal in the
spatial domain is optimised.  However, no sampling theorem on the
sphere reaches the optimal number of samples in the spatial domain
suggested by the $\elmax^2$ dimensionality of the signal in the
harmonic domain.  Consequently, the dimensionality of the problem is
reduced by recovering the harmonic coefficients $\vect{\hat{x}}$
directly:
\begin{equation}
\label{eqn:recon_harmonic}
\vect{\hat{x}}^\star =
\underset{\vect{\hat{x}}}{\arg \min} \:
\| \sparmat \vect{\hat{x}} \|_{\rm TV} \:\: \mbox{such that} \:\:
\| \vect{y} - \sensmat \sparmat \vect{\hat{x}}\|_2 \leq \epsilon
\spcend ,
\end{equation}
where $\sparmat\in\complex^{\N\times\elmax^2}$ represents the inverse
spherical harmonic transform; the signal on the sphere is recovered by
$\vect{x}^\star = \sparmat \vect{\hat{x}}^\star$.  For this problem
the dimensionality of the signal directly recovered $\vect{\hat{x}}$
is identical for both sampling theorems, however sparsity in the
spatial domain remains superior (\ie\ fewer non-zero values) for the
MW sampling theorem.

\begin{figure}
\centering
\includegraphics[width=110mm]{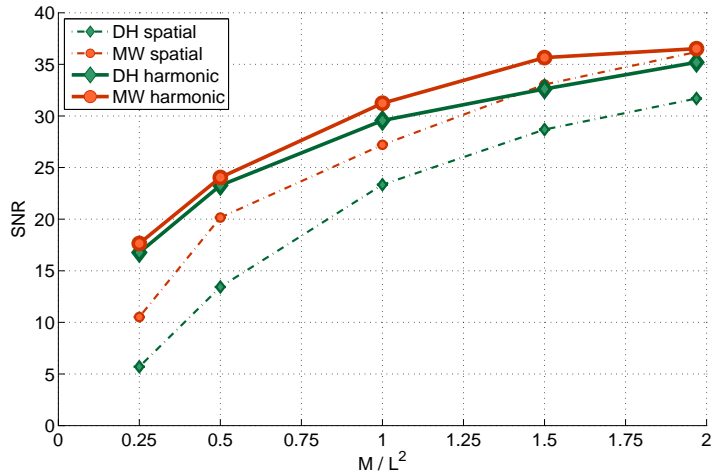}
\caption{Reconstruction performance for the DH and MW sampling theorems}
\label{fig:snr_vs_m}
\end{figure}

Reconstruction performance is plotted in \fig{\ref{fig:snr_vs_m}} when
solving the inpainting problem in the spatial and harmonic domains,
through \eqn{\ref{eqn:recon_spatial}} and
\eqn{\ref{eqn:recon_harmonic}} respectively, for both sampling
theorems (averaged over ten simulations of random measurement
operators and independent and identically distributed Gaussian noise).
Strictly speaking, compressed sensing corresponds to the measurement ratio
$\nmeas/\elmax^2<1$ when considering the harmonic representation of
the signal.  Nevertheless, experiments are extended to $\nmeas/\elmax^2
\sim 2$, corresponding to the equivalent of complete sampling on the
MW grid.
When solving the inpainting problem in the spatial domain we see a
large improvement in reconstruction quality for the MW sampling
theorem when compared to the DH sampling theorem.  This is due to the
enhancement in both dimensionality and sparsity afforded by the MW
sampling theorem in this setting.  When solving the inpainting problem
in the harmonic domain, we see a considerable improvement in
reconstruction quality for each sampling theorem, since the
dimensionality of the recovered signal is optimal in harmonic space.
For harmonic reconstructions, the MW sampling theorem remains superior
to the DH sampling theorem due to the enhancement in sparsity (but not
dimensionality) that it affords in this setting. In all cases, the
superior performance of the MW sampling theorem is evident.
In \fig{\ref{fig:spheres}} example reconstructions are shown, where
the superior quality of the MW reconstruction is again clear.

\section{CONCLUSIONS} 
\label{sec:conclusions}

Although compressive sensing states that sparse or compressible
signals may be acquired with fewer samples than standard sampling
theorems would suggest, the sampling theorem adopted nevertheless has
an important influence on the performance of compressive sensing
reconstruction.  In Euclidean space, Shannon's sampling theorem
provides an optimal sampling for regular grids, leading to a unique
sampling theorem.  On the sphere, however, no sampling theorem is
optimal, with different sampling theorems requiring a differing number
of samples.  The MW sampling theorem\cite{mcewen:fssht} has been
developed only recently and achieves a more efficient sampling of the
sphere than alternatives, requiring fewer than half as many samples as
the canonical DH sampling theorem\cite{driscoll:1994}, while still
capturing all of the information content of a band-limited signal.  A
reduction by a factor of two in the number of samples between the DH
and MW sampling theorems has important implications for compressive
sensing on the sphere, both in terms of the dimensionality and
sparsity of signals.  The more efficient sampling of the MW sampling
theorem has been shown to enhance the performance of compressed
sensing reconstruction on the sphere, as illustrated with an
inpainting problem.\cite{mcewen:css2}


\acknowledgments     
 
JDM is supported by the Swiss National Science Foundation (SNSF) under
grant 200021-130359.  YW is supported by the Center for
Biomedical Imaging (CIBM) of the Geneva and Lausanne Universities,
EPFL, and the Leenaards and Louis-Jeantet foundations, and by
the SNSF under grant PP00P2-123438.


\bibliography{bib}   
\bibliographystyle{spiebib}   

\end{document}